\documentclass[10pt]{iopart}

\usepackage{iopams}
\usepackage{graphicx} 
\usepackage{dcolumn}
\usepackage{upgreek }
\usepackage{times}
\usepackage{textcomp}
\usepackage{ wasysym }
\usepackage{titlesec}
\usepackage{float}
\usepackage{color,xcolor}
\usepackage{graphicx,times} 
\usepackage{cuted} \stripsep -3pt plus 3pt minus 2pt
\usepackage{epstopdf}
\usepackage{graphicx}
\usepackage{subfigure}
\usepackage{makecell}
\usepackage{multirow}
\renewcommand{\figurename}{Figure}
\usepackage[colorlinks,linkcolor=blue, urlcolor=blue, anchorcolor=blue, citecolor=blue]{hyperref}
\usepackage{longtable,booktabs}
\usepackage{fouriernc}
\usepackage{epsfig,graphics,graphicx}
\usepackage{bm}
\usepackage{color}
\usepackage{flushend}
\usepackage{ulem}
\usepackage[sort&compress,square,numbers]{natbib}

\begin{document}
\title[Multi-Fluid Equilibrium of Proton-Boron Spherical Tokamaks]{Development of a Reduced Multi-Fluid Equilibrium Model and Its Application to Proton-Boron Spherical Tokamaks}

\author{
Huasheng Xie$^{1,*}$, Xingyu Li$^{2}$, Jiaqi Dong$^{1,3}$, Zhiwei Ma$^{4}$, Yunfeng Liang$^{5,1}$, Yuejiang Shi$^{1}$, Wenjun Liu$^{1}$, Yueng-Kay Martin Peng$^{1}$, Lai Wei$^{2}$, Zhengxiong Wang$^{2}$ and Hanyue Zhao$^{1}$
}

\address{

$^1$ ENN Science and Technology Development Co., Ltd., Langfang 065001, People's Republic of China

$^2$ School of Physics, Dalian University of Technology, Dalian 116024, People’s Republic of China

$^3$  Southwestern Institute of Physics, Chengdu 610225, People’s Republic of China

$^4$ Institute for Fusion Theory and Simulation, Zhejiang University, Hangzhou 310027, People’s Republic of
China

$^5$ Forschungszentrum Jülich GmbH, Institute of Fusion Energy and Nuclear Waste Management Plasmaphysik, Jülich 52425, Germany
}
\eads{\mailto{huashengxie@gmail.com, xiehuasheng@enn.cn}}

\begin{indented}
\item[\today]
\end{indented}

\begin{abstract}

Proton-Boron (p-$^{11}$B) fusion represents a promising pathway toward aneutronic clean energy but requires extremely high ion temperatures and robust magnetic confinement. Spherical Tokamaks/Torus (ST) driven by high-power neutral beam injection are a primary candidate for this regime. In such devices, the combination of strong toroidal rotation and the significant mass disparity between protons and boron ions leads to complex multi-fluid effects---specifically centrifugal species separation and electrostatic polarization---which standard single-fluid magnetohydrodynamic (MHD) models fail to capture. Conversely, comprehensive multi-fluid models that include poloidal flows often suffer from numerical stiffness and excessive complexity, hindering their use in routine engineering analysis.  To address these challenges, we have developed a \textit{reduced multi-fluid equilibrium model} designed to balance physical fidelity with computational robustness. By retaining the dominant toroidal rotation and self-consistent electrostatic potential while neglecting secondary effects such as poloidal flow inertia and pressure anisotropy, the model is formulated as a generalized Grad-Shafranov equation coupled with species-specific Bernoulli relations and a quasi-neutrality constraint. The model is applied to analyze the equilibrium configurations of two representative p-$^{11}$B ST devices designed by the ENN Group: the experimental EHL-2 and the reactor-scale EHL-3B. Simulation results demonstrate that the equilibrium modification is governed by the ion Mach number ($M$). In the low-rotation regime ($M < 0.5$), multi-fluid effects are weak, and the solution converges toward the single-fluid limit. However, in the high-rotation regime ($M > 2$), strong centrifugal forces drive significant boron accumulation at the low-field side (LFS) and generate an internal electrostatic potential on the order of 10 kV. These results confirm the necessity of multi-fluid modeling for accurate p-$^{11}$B reactor design and provide a robust theoretical foundation for future investigations into stability, transport, and free-boundary dynamics.
\end{abstract}

\maketitle
\ioptwocol

\section{Introduction}\label{sec:intro}

Proton-Boron (p-$^{11}$B) fusion has recently re-emerged as a high-interest research area due to its potential for aneutronic energy production, which minimizes neutron-induced damage and radioactive waste \cite{Liu2024}. To achieve viable fusion gain, ion temperatures ($T_i$) must exceed 100 keV, a regime far beyond that of conventional Deuterium-Tritium (D-T) reactors. Spherical Tokamaks/Torus (ST) are uniquely suited for this mission due to their ability to achieve high volume-averaged beta ($\beta$) and good confinement properties in a compact geometry. In these high-performance regimes, strong Neutral Beam Injection (NBI) is typically required to maintain both the thermal energy and the plasma current, which inherently drives significant toroidal rotation \cite{Liang2025}.

Historically, the single-fluid Grad-Shafranov (GS) model has been the workhorse of tokamak equilibrium analysis \cite{Freidberg1987, Jardin2010, Pereverzev2002}. However, p-$^{11}$B plasmas are intrinsically multi-species environments where the mass disparity between protons ($1 \, m_p$) and boron ions ($\sim 11 \, m_p$) is extreme. Under strong toroidal rotation, the centrifugal force scales linearly with mass, causing heavy boron ions to shift toward the low-field side (LFS) of magnetic surfaces. This spatial separation induces charge polarization, necessitating a self-consistent electrostatic potential ($\Phi$) to preserve quasi-neutrality. Conventional single-fluid MHD models, which assume constant pressure on flux surfaces or a collective single-fluid centrifugal potential \cite{Hameiri1983, Chen2022, Feng2024, Li2026}, fundamentally lack the degrees of freedom to describe this differential species transport and the resulting potential structure.

While comprehensive single-fluid and multi-fluid models incorporating poloidal flows and pressure anisotropy exist \cite{Jardin2010, Qu2014,Guazzotto2004, Guazzotto2015, Steinhauer1999, Qerushi2003, Galeotti2011, Ishida2010, Ishida2012, Ishida2015, Ishida2020,Iacono1990,Thyagaraja2006,Hole2009,Evangelias2016}, they are often characterized by significant mathematical complexity. For instance, the inclusion of poloidal flow inertia introduces trans-sonic singularities where the governing equations change type from elliptic to hyperbolic, leading to severe numerical stiffness. These issues often hinder the use of such models as standard baseline tools for routine reactor design and parameter scanning. There is, therefore, a clear need for a "reduced" multi-fluid model that captures the primary physics of species separation and potential formation while maintaining the robustness and efficiency of the traditional GS framework.

The primary objective of this paper is to establish such a foundational reduced multi-fluid equilibrium model for p-$^{11}$B spherical tokamaks. Based on the physical characteristics of STs heated by NBI, we adopt a strategic simplification: we prioritize toroidal inertia and electrostatic coupling while deferring higher-order complexities like poloidal flows and pressure anisotropy to subsequent work. This provides a tractable theoretical baseline suitable for engineering optimization. We demonstrate the utility of this model by investigating two representative ENN ST devices \cite{Liu2024}: the experimental EHL-2 and the reactor-scale EHL-3B. 

The remainder of this paper is organized as follows. Section \ref{sec:model} formulates the mathematical model, including the species-specific Bernoulli relations and the generalized Grad-Shafranov equation. Section \ref{sec:numerical} describes the numerical algorithm and the feedback mechanism for the fixed current constraint. Section \ref{sec:application} presents a comparative analysis of the EHL-2 and EHL-3B equilibria. Finally, a summary and discussion are provided in Section \ref{sec:summary}.

\begin{figure*}[htbp]
\centering
\includegraphics[width=17.0cm]{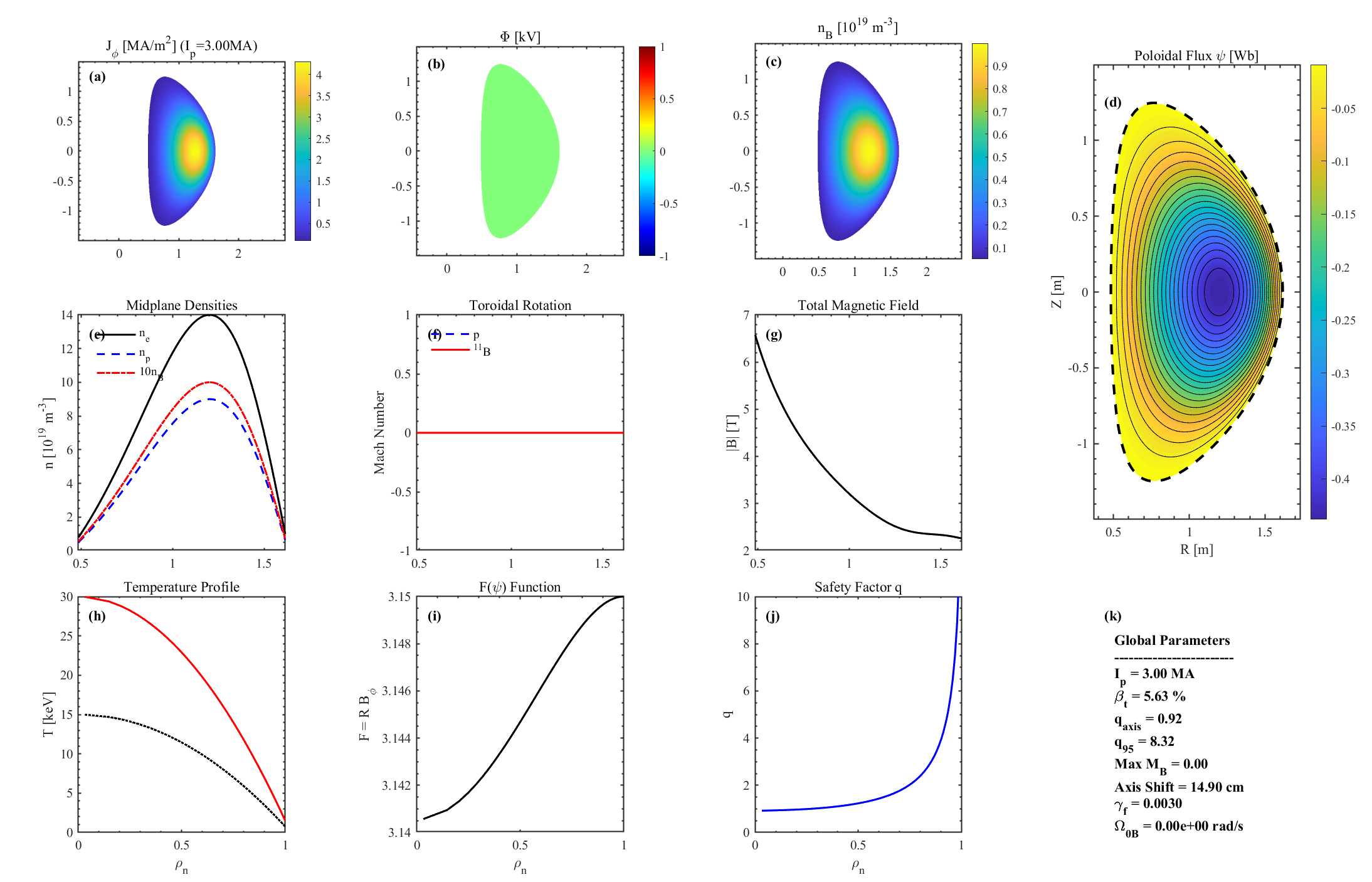}
\caption{Multi-fluid equilibrium analysis of the EHL-2 configuration in the static limit ($M=0$). The panels display: (a) Toroidal current density $J_\phi$; (b) Electrostatic potential $\Phi$ (zero in static case); (c) Boron density $n_B$; (d) Poloidal magnetic flux $\psi$; (e) Midplane density profiles at $Z=0$; (f) Toroidal Mach number (zero); (g) Total magnetic field $B_{tot}$; (h)-(j) Flux-coordinate profiles of Temperature $T(\psi)$, Poloidal current function $F(\psi)$, and Safety factor $q(\psi)$; (k) Global parameters table. Note that at $M=0$, all species distributions are uniform on flux surfaces, and the solution recovers the standard single-fluid MHD result.}
\label{fig:EHL2_w0}
\end{figure*}

\begin{figure*}[htbp]
\centering
\includegraphics[width=17.0cm]{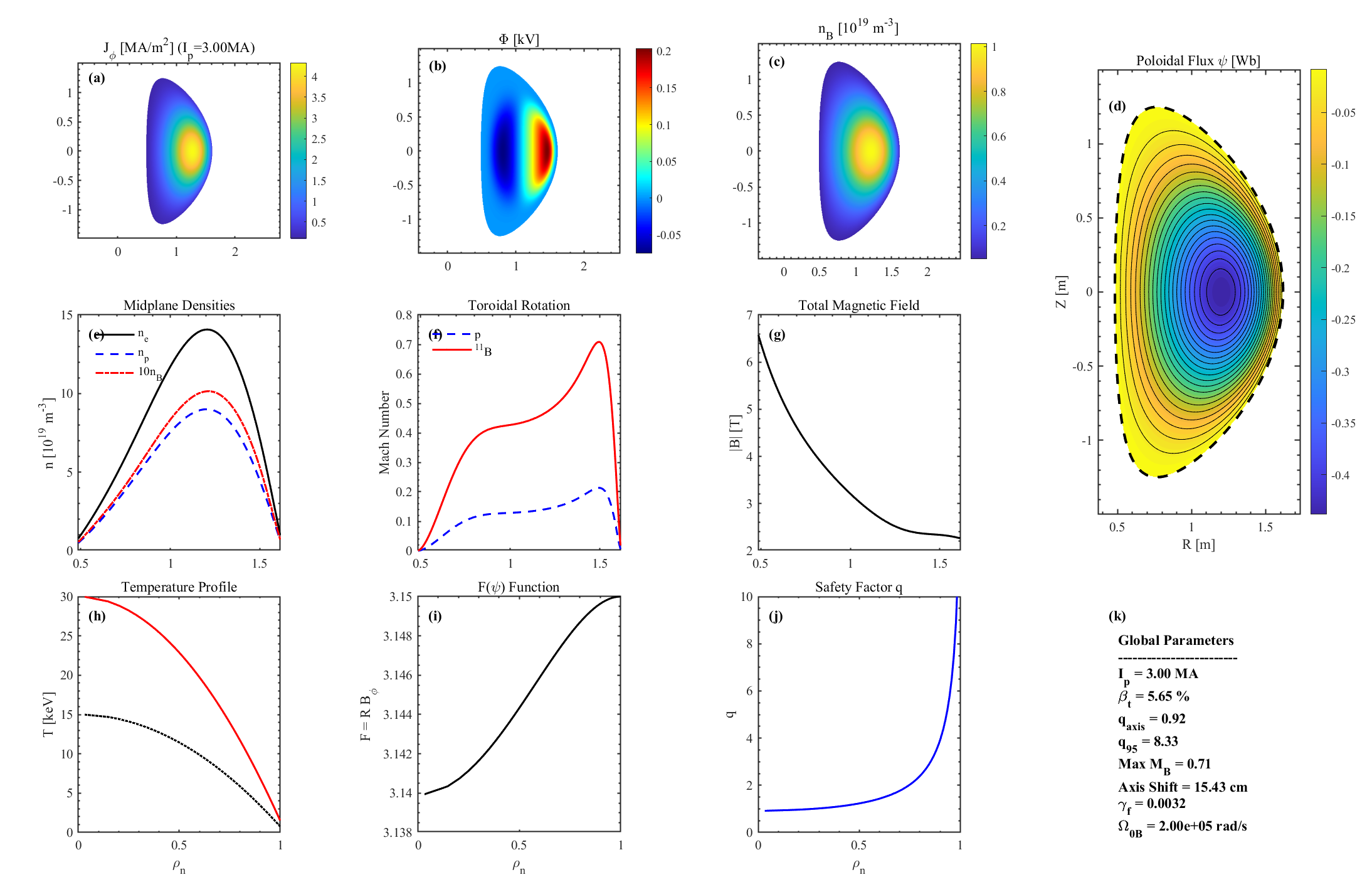}
\caption{EHL-2 equilibrium with moderate toroidal rotation ($M_B \sim 0.5$). Small centrifugal modifications are observed in the current density and pressure distributions, while the boron density $n_B$ begins to show a slight outboard shift. The self-consistent electrostatic potential $\Phi$ remains below $0.2$~kV, indicating that single-fluid approximations are still relatively accurate in this low-Mach regime.}
\label{fig:EHL2_w2}
\end{figure*}

\begin{figure*}[htbp]
\centering
\includegraphics[width=17.0cm]{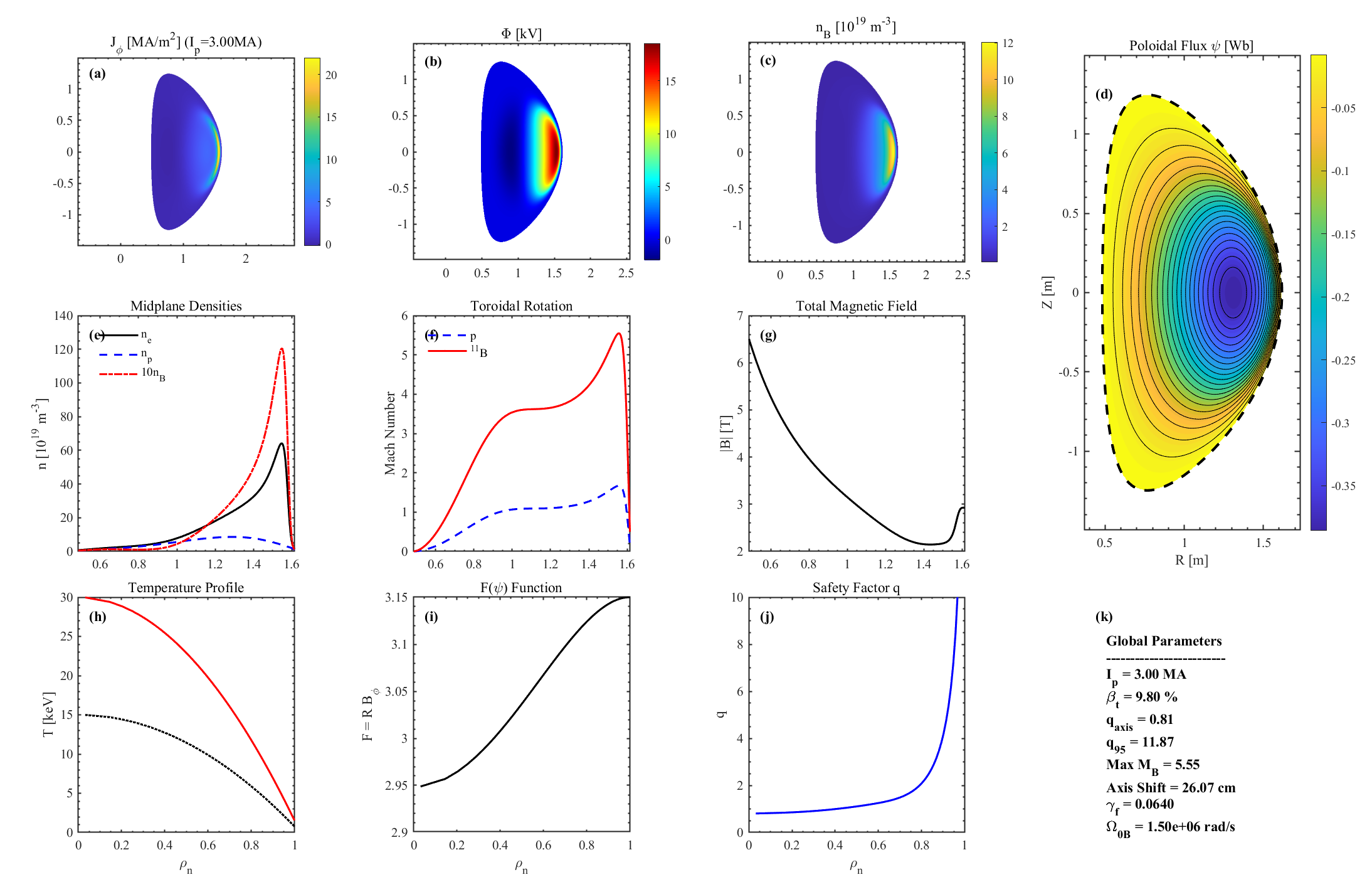}
\caption{EHL-2 equilibrium under extreme toroidal rotation ($M_B \sim 5$, corresponding to $u_\phi \approx 1500$ km/s)\cite{Liang2025}. This case tests the numerical robustness of the model in a high-Mach regime. (a) The toroidal current density $J_\phi$ becomes highly hollow and localized at the LFS. (c) The boron density $n_B$ is entirely expelled to the outboard edge, exhibit an extreme "crescent-shaped" distribution, leaving the core depleted. (b) A massive electrostatic potential $\Phi \approx 10$ kV is generated to shield electrons from the centrifugal force. Note that while $1500$ km/s is an extreme theoretical limit for EHL-2, it demonstrates the model's capability to capture strong shock-free separation physics.}
\label{fig:EHL2_w15}
\end{figure*}

\begin{figure*}[htbp]
\centering
\includegraphics[width=17.0cm]{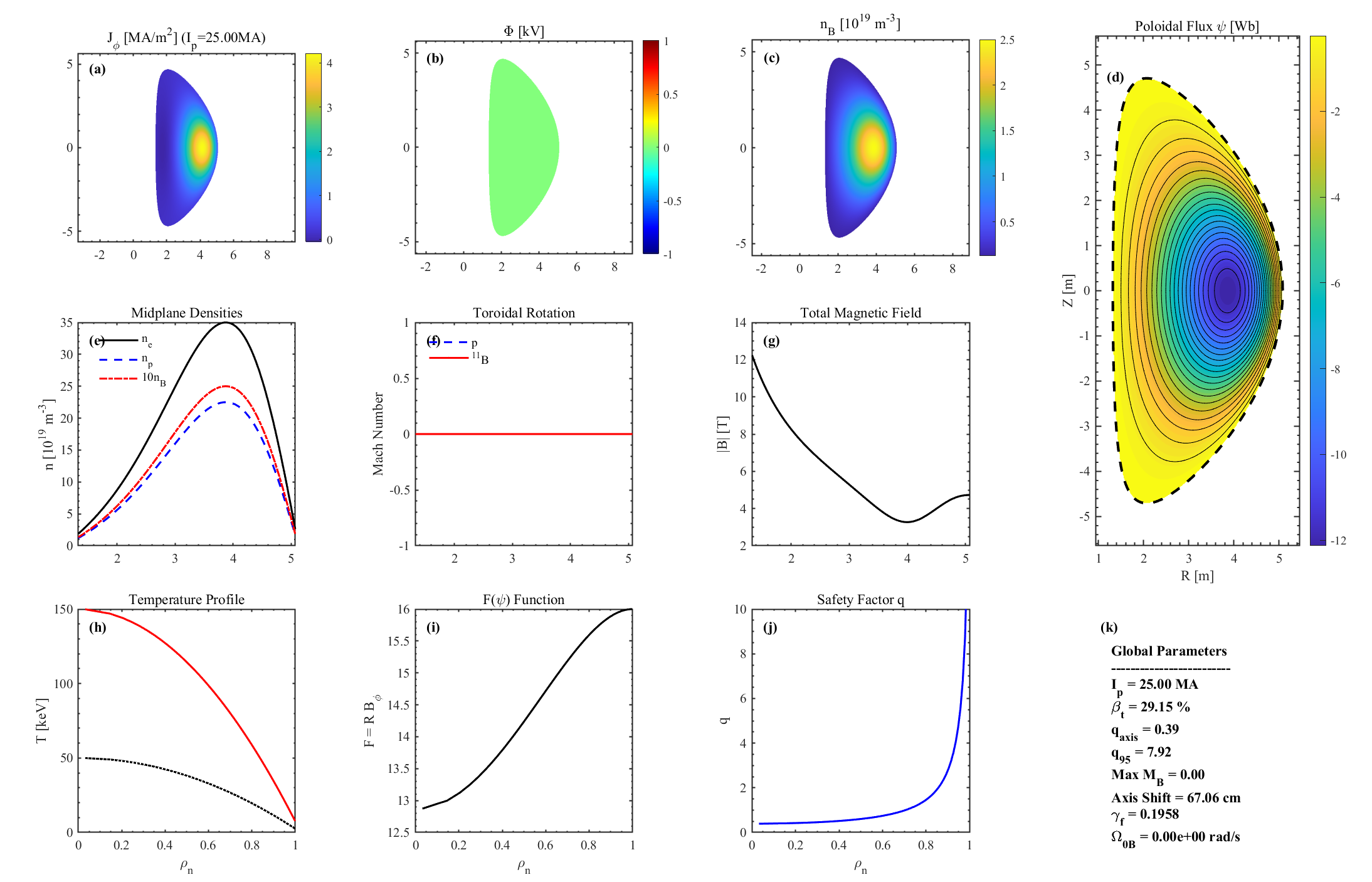}
\caption{Baseline multi-fluid equilibrium for the reactor-scale EHL-3B device in the static case ($M=0$). Despite the high plasma beta ($\beta_t \approx 30\%$) and large Shafranov shift ($\sim 67$ cm), the species densities ($n_e, n_p, n_B$) remain perfectly aligned with the magnetic flux surfaces $\psi$ in the absence of rotation.}
\label{fig:EHL3B_w0}
\end{figure*}

\begin{figure*}[htbp]
\centering
\includegraphics[width=17.0cm]{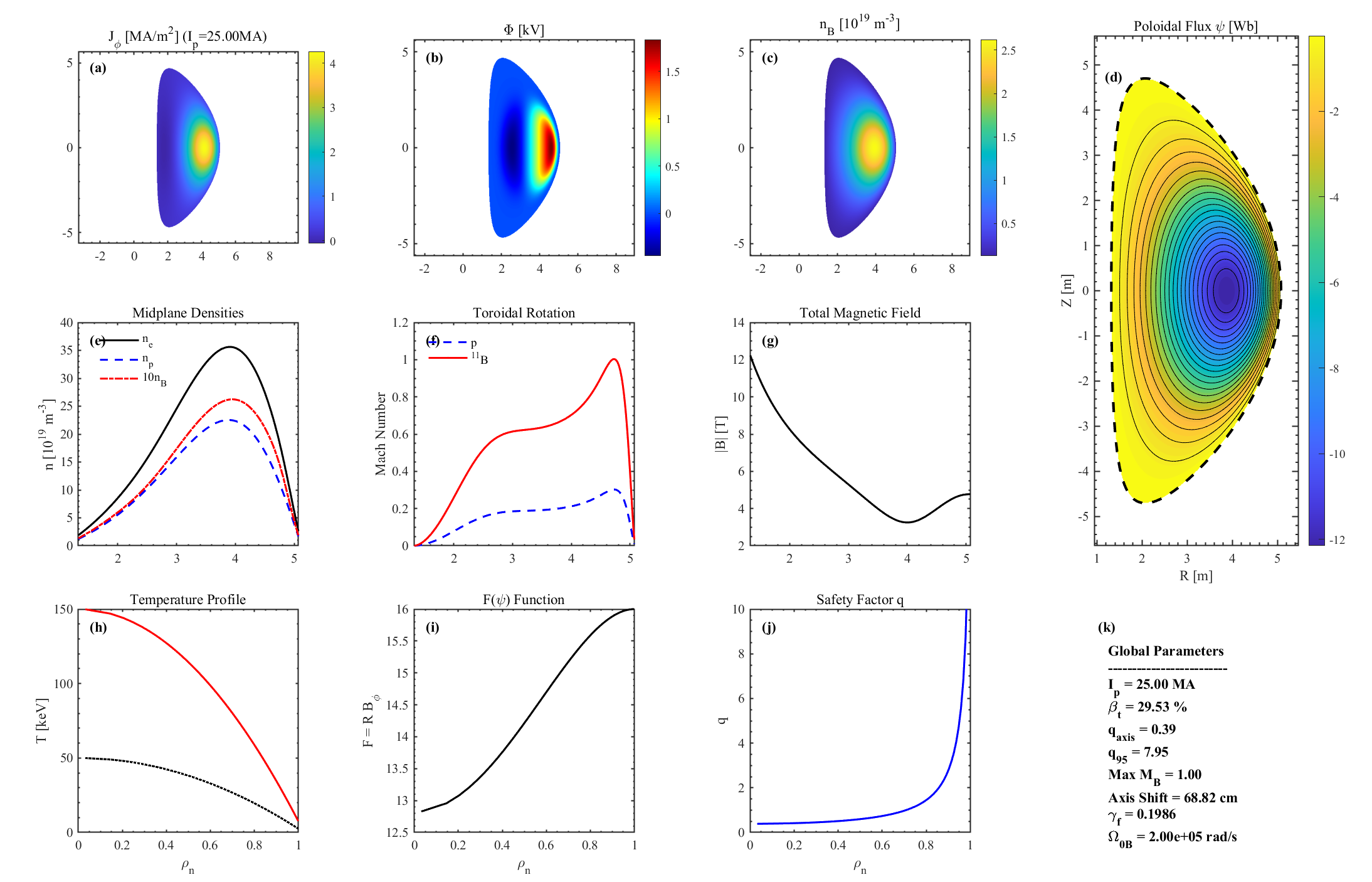}
\caption{EHL-3B equilibrium with moderate rotation ($M_B \sim 1.0$). At this reactor scale, the absolute rotation frequency is lower than in EHL-2 to achieve the same Mach number due to the larger radius. The centrifugal separation is clearly visible in the midplane density profiles (Panel e), and an electrostatic potential of around 1 kV is generated to maintain quasi-neutrality between protons and boron ions.}
\label{fig:EHL3B_w2}
\end{figure*}

\begin{figure*}[htbp]
\centering
\includegraphics[width=17.0cm]{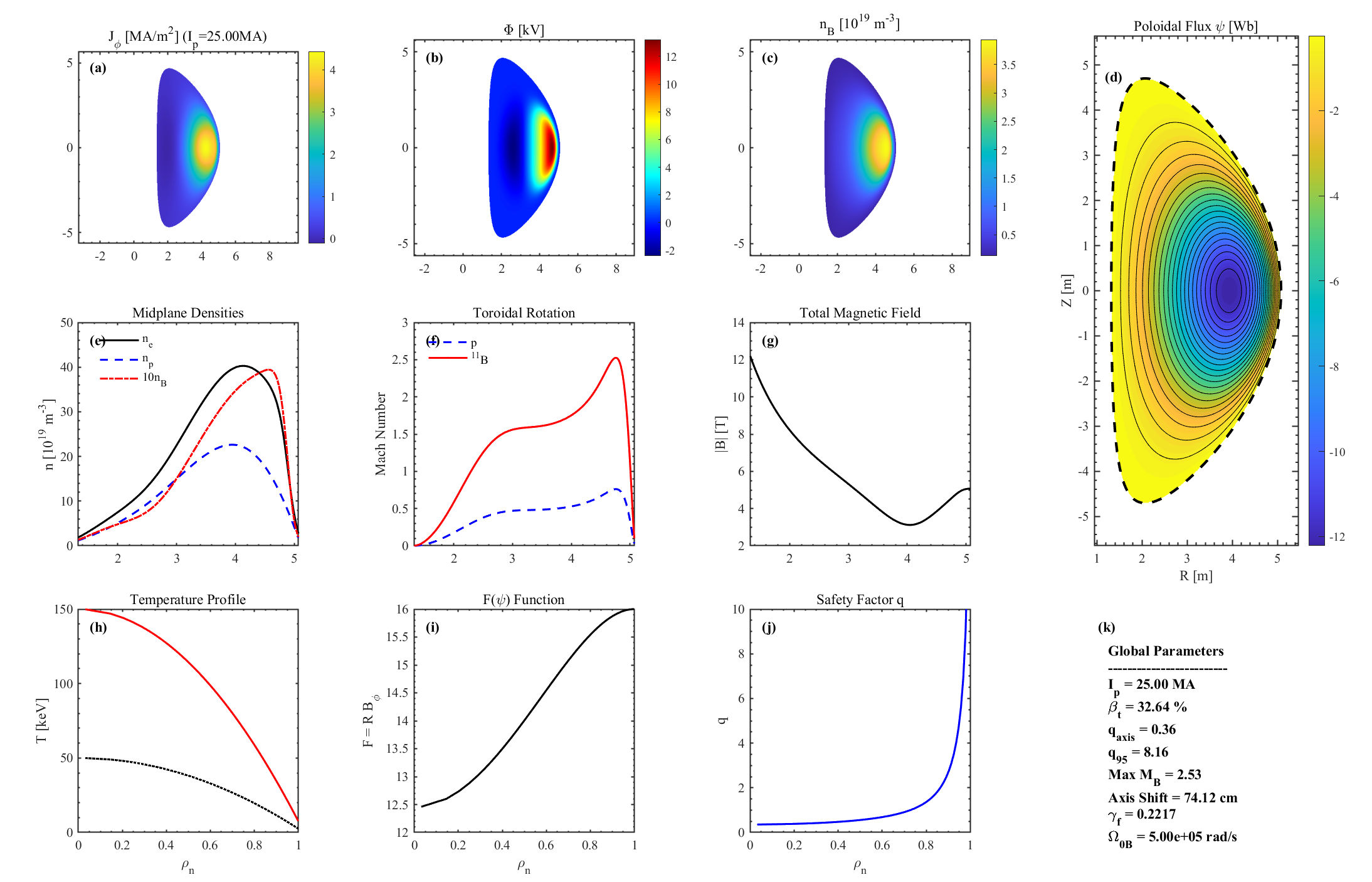}
\caption{EHL-3B equilibrium in the high-rotation reactor regime ($M_B \sim 2$). The heavy $^{11}$B ions are strongly localized at the LFS outboard midplane  (Panel c). The self-consistent potential $\Phi$ reaches a peak value of approximately $10$~kV. The resulting spatial mismatch between proton and boron densities (Panel e) highlights the crucial need for multi-fluid modeling in predicting fusion power and stability for $p\mbox{-}^{11}$B burning plasmas.}
\label{fig:EHL3B_w5}
\end{figure*}

\section{Mathematical Formulation}\label{sec:model}

The core philosophy of the proposed model is to capture the leading-order physics of species separation and electrostatic coupling while maintaining numerical robustness. We establish a system where the magnetic geometry and the particle distributions are self-consistently coupled through the electrostatic potential.

\subsection{Governing Equations and Assumptions}

We consider an axisymmetric plasma consisting of multiple species indexed by $s$, including protons ($p$), boron ions ($^{11}$B), and electrons ($e$). The equilibrium is governed by the steady-state multi-fluid momentum equation for each species $s$:
\begin{equation}
    m_s n_s (\mathbf{u}_s \cdot \nabla) \mathbf{u}_s = -\nabla p_s + n_s q_s (\mathbf{E} + \mathbf{u}_s \times \mathbf{B}),
    \label{eq:momentum}
\end{equation}
where $m_s, n_s, \mathbf{u}_s, p_s, q_s$ represent the mass, density, velocity, pressure, and charge of species $s$, respectively. To achieve a "reduced" yet physically consistent model suitable for p-11B ST design, the following assumptions are adopted:
\begin{enumerate}
    \item \textbf{Axisymmetry:} $\partial/\partial \phi = 0$ for all physical quantities. The magnetic field is expressed as $\mathbf{B} = \nabla \phi \times \nabla \psi + F(\psi) \nabla \phi$, where $\psi$ is the poloidal magnetic flux and $F(\psi) = R B_\phi$ is the poloidal current function.
    \item \textbf{Dominant Toroidal Rotation:} The flow is assumed to be purely toroidal, $\mathbf{u}_s = R \Omega_s(\psi) \hat{\phi}$, for the purpose of inertia calculation. Poloidal flow inertia is neglected to avoid trans-sonic singularities.
    \item \textbf{Isothermal Surfaces:} The temperature of each species is assumed to be a flux function, $T_s = T_s(\psi)$, due to high parallel thermal conductivity. The pressure is isotropic, $p_s = n_s k_B T_s$.
    \item \textbf{Massless Electrons:} Since $m_e \ll m_i$, electron inertia is neglected, so their centrifugal potential is set to zero ($W_e^{cf} = 0$).
\end{enumerate}

\subsection{Physical Justification of the Reduced Model}

The "reduced" nature of this model stems from two primary physical simplifications: the neglect of poloidal flow inertia and the assumption of isotropic pressure. 

First, in high-temperature tokamaks, poloidal rotation is strongly damped by neoclassical magnetic pumping on a timescale much faster than toroidal momentum confinement \cite{Hinton1985}. Consequently, the poloidal Mach number typically remains small ($M_p \ll 1$). By neglecting poloidal inertia, the Grad-Shafranov operator remains strictly elliptic, avoiding the mathematical singularities associated with the poloidal sonic resonance found in full two-fluid models \cite{Guazzotto2015}.

Second, while intense NBI can induce pressure anisotropy ($p_\parallel \neq p_\perp$), the resulting equilibrium modification is generally a correction of the order of the plasma beta ($\beta$). In contrast, for high-mass species like boron ($m_B \approx 11 m_p$), the centrifugal force scales with the square of the Mach number ($M^2$). In the strongly rotating regimes intended for p-$^{11}$B STs ($M_B > 1$), the spatial redistribution of ions driven by the centrifugal potential is the dominant effect. Thus, the isotropic pressure assumption provides a reliable baseline for studying species separation.

\subsection{Species Density and Bernoulli Relations}

Projecting Eq.~(\ref{eq:momentum}) along the magnetic field $\mathbf{B}$ and integrating along the field line leads to the generalized Boltzmann (Bernoulli) distribution for each species:
\begin{equation}
    n_s(R, \psi) = N_s(\psi) \exp \left( \frac{W_s^{cf}(R, \psi) - q_s \Phi(R, \psi)}{k_B T_s(\psi)} \right),
    \label{eq:density}
\end{equation}
where $N_s(\psi)$ is the reference density profile defined at a reference radius $R_0$ (typically the geometric center), and $W_s^{cf}$ is the centrifugal potential energy:
\begin{equation}
    W_s^{cf}(R, \psi) = \frac{1}{2} m_s \Omega_s^2(\psi) (R^2 - R_0^2).
\end{equation}
Equation (\ref{eq:density}) explicitly shows that heavier species are localized to the low-field side (LFS) due to $W_s^{cf}$, while the electrostatic potential $\Phi$ adjusts to maintain charge balance.

\subsection{Quasineutrality Closure}

The self-consistent electrostatic potential $\Phi(R, Z)$ is determined by the quasi-neutrality condition:
\begin{equation}
    \sum_s q_s n_s(R, \psi, \Phi) = 0.
    \label{eq:neutrality}
\end{equation}
Substituting Eq.~(\ref{eq:density}) into Eq.~(\ref{eq:neutrality}) results in a nonlinear algebraic equation for $\Phi$ at each spatial point $(R, Z)$. Unlike single-fluid models where pressure is a flux function, here the local density $n_s$ depends on both $R$ and $\Phi$. Consequently, the total pressure $P_{\rm tot} = \sum_s n_s k_B T_s$ becomes a function of $(R, \psi)$.

\subsection{Generalized Grad-Shafranov Equation}

The force balance across magnetic flux surfaces leads to the generalized Grad-Shafranov (GS) equation:
\begin{equation}
    \Delta^* \psi =\mu_0RJ_{\phi},~J_{\phi}= -R \left( \frac{\partial P_{\rm tot}}{\partial \psi} \right)_R - \frac{1}{\mu_0 R}F \frac{dF}{d\psi},
    \label{eq:GS}
\end{equation}
where $\Delta^* = R \frac{\partial}{\partial R} (\frac{1}{R} \frac{\partial}{\partial R}) + \frac{\partial^2}{\partial Z^2}$ is the elliptic operator. The source term $(\partial P_{\rm tot} / \partial \psi)_R$ must be evaluated using the chain rule to account for the implicit dependence of $\Phi$ on $\psi$.

Defining the auxiliary variable $A_s$ for each species as:
\begin{equation}
    A_s = \frac{N_s'}{N_s} - \frac{T_s'}{T_s} \left( \frac{W_s^{cf} - q_s \Phi}{k_B T_s} \right) + \frac{m_s \Omega_s \Omega_s' (R^2 - R_0^2)}{k_B T_s},
    \label{eq:As_def}
\end{equation}
where the prime ($'$) denotes $d/d\psi$. Differentiating Eq.~(\ref{eq:neutrality}) with respect to $\psi$ allows us to solve for the gradient of the potential:
\begin{equation}
    \left( \frac{\partial \Phi}{\partial \psi} \right)_R = \frac{\sum_s q_s n_s A_s}{\sum_s q_s^2 n_s / (k_B T_s)}.
    \label{eq:dPhi_dpsi}
\end{equation}
Consequently, the species density gradient is given by:
\begin{equation}
    \left( \frac{\partial n_s}{\partial \psi} \right)_R = n_s \left[ A_s - \frac{q_s}{k_B T_s} \left( \frac{\partial \Phi}{\partial \psi} \right)_R \right].
    \label{eq:dns_dpsi}
\end{equation}
Finally, the total pressure gradient source term in Eq. (\ref{eq:GS}) is assembled as:
\begin{equation}
    \left( \frac{\partial P_{\rm tot}}{\partial \psi} \right)_R = \sum_s \left[ k_B T_s \left( \frac{\partial n_s}{\partial \psi} \right)_R + n_s k_B T_s' \right].
    \label{eq:dPtot_dpsi}
\end{equation}
Equations (\ref{eq:density})--(\ref{eq:GS}) with (\ref{eq:As_def})--(\ref{eq:dPtot_dpsi}) constitute a closed system that describes the multi-fluid equilibrium with self-consistent centrifugal separation and electrostatic polarization.

\subsection{Model Comparison and Rationale}

The proposed model represents a strategic trade-off between standard MHD and comprehensive multi-fluid theories. A primary difference from the frameworks of Steinhauer \cite{Steinhauer1999} and Galeotti \cite{Galeotti2011} is our choice of the invariant surface function. While they define profiles on drift surfaces $Y_s = \psi + m_s R u_{\phi, s} / q_s$, we assume dependence on the magnetic flux $\psi$ directly. This simplification neglects finite-orbit-width corrections but ensures the Grad-Shafranov operator remains elliptic and avoids an excessive number of free parameters, which often makes comprehensive models numerically stiff and difficult to constrain \cite{Ishida2012}.

Compared to Ishida’s models \cite{Ishida2010, Ishida2012, Ishida2015, Ishida2020}, which incorporate poloidal ion flows, relativistic electrons, and energetic particles, our reduced approach focuses on dominant toroidal rotation and species separation. Notably, we retain the macroscopic effect of electron poloidal flow to determine the toroidal flux function $F(\psi)$ self-consistently. Consequently, this model serves as a natural, computationally efficient extension of single-fluid toroidal rotation models tailored for the design and analysis of high-performance p-$^{11}$B spherical torus devices.

\subsection{Profile Functions and Normalization}

To close the system, we specify profile functions as a function of the normalized poloidal flux $\psi_n = (\psi - \psi_{\rm axis})/(\psi_{\rm edge} - \psi_{\rm axis})$, with also $\rho_n=\sqrt{\psi_n}$. The following parameterized forms are used:

\noindent\textbf{Reference Density:}
\begin{equation}
    N_s(\psi_n) = (N_{s0}- N_{s,\rm edge})  \left(1 - \psi_n^{\beta_{ns}}\right)^{\alpha_{ns}} + N_{s,\rm edge} ,
\end{equation}
\noindent\textbf{Temperature:}
\begin{equation}
    T_s(\psi_n) = (T_{s0}- T_{s,\rm edge}) \left(1 - \psi_n^{\beta_{Ts}}\right)^{\alpha_{Ts}} + T_{s,\rm edge},
\end{equation}
\noindent\textbf{Rotation Frequency:}
\begin{equation}
    \Omega_s(\psi_n) = \Omega_{s0} \left(1 - \psi_n^{\beta_{\Omega s}}\right)^{\alpha_{\Omega s}},
\end{equation}
where $\Omega_{s0}$ is the rotation frequency at the magnetic axis. 
The corresponding thermal Mach number for species $s$ is defined as:
\begin{equation}
    M_s(\psi_n) = \frac{R_{0} \, \Omega_s(\psi_n)}{v_{th,s}} = \frac{R_{0} \, \Omega_s(\psi_n)}{\sqrt{k_B T_s(\psi_n) / m_s}}.
    \label{eq:Mach_def}
\end{equation}
The toroidal field function is parameterized to allow for feedback control of the total current $I_p$:
\begin{equation}
    F(\psi_n) = F_{\rm vac} \left[ 1 - \gamma_f (1 - \psi_n^{\beta_f})^{\alpha_f} \right],
    \label{eq:F_function}
\end{equation}
where $F_{\rm vac} = R_0 B_0$ is the vacuum toroidal field constant, and $\gamma_f$ is the free parameter adjusted to match the target $I_p=\int J_\phi dRdZ$. These functions provide the $7+1=8$ input degrees of freedom for the model.

\section{Numerical Algorithm}\label{sec:numerical}

The coupled system of equations derived in Section \ref{sec:model}---consisting of the generalized Grad-Shafranov (GS) equation, species-specific Bernoulli relations, and the quasi-neutrality constraint---poses a highly nonlinear problem. To solve this efficiently, we have developed a numerical code, \textsc{Nfeq} (N-Fluid EQuilibrium), employing a nested iterative scheme that ensures self-consistency between the magnetic geometry and the multi-species particle distributions.

\subsection{Nested Iterative Scheme}

The problem is discretized on a rectangular $(R, Z)$ mesh using the Finite Difference Method (FDM). For the fixed-boundary cases studied here, the plasma boundary is prescribed. A masking technique is used to distinguish the plasma domain from the vacuum region. 

The overall solver follows a Picard iteration loop (outer loop) for the poloidal flux $\psi$, within which a Newton-Raphson solver (inner loop) determines the electrostatic potential $\Phi$:

\begin{enumerate}
    \item \textbf{Initialization:} An initial guess $\psi^{(0)}(R, Z)$ is provided. The normalized flux $\psi_n$ and input profile functions (density, temperature, rotation) are computed.
    
    \item \textbf{Inner Loop (Potential Solver):} At each grid point $(R, Z)$ within the plasma, the quasi-neutrality condition (Eq.~\ref{eq:neutrality}) is solved for $\Phi^{(k)}(R, Z)$. Given the algebraic and local nature of this constraint, we utilize a vectorized Newton-Raphson method. The analytic Jacobian, $\partial (\sum_s q_s n_s) / \partial \Phi = -\sum_s q_s^2 n_s / (k_B T_s)$, is used to ensure quadratic convergence.
    
    \item \textbf{Source Term Assembly:} Once $\Phi^{(k)}$ and $n_s^{(k)}$ are converged, the auxiliary variables $A_s$ and the potential gradient $(\partial \Phi / \partial \psi)_R$ are calculated. These are used to construct the total pressure gradient $(\partial P_{\rm tot} / \partial \psi)_R$ via Eq.~(\ref{eq:dPtot_dpsi}).
    
    \item \textbf{Field Solve:} The toroidal current density $J_\phi(R, Z)$ is assembled. The linear elliptic equation $\Delta^* \psi^* = \mu_0 R J_\phi$ is solved using a direct sparse solver to obtain a provisional flux $\psi^*$.
    
    \item \textbf{Relaxation:} To manage the strong nonlinearity introduced by the exponential centrifugal terms, an under-relaxation scheme is applied: $\psi^{(k+1)} = (1 - \omega) \psi^{(k)} + \omega \psi^*$. The relaxation factor $\omega$ is typically set between $0.05$ and $0.5$ for high-Mach number cases to maintain numerical stability.
\end{enumerate}

\subsection{Feedback Control for Fixed Total Current}

In multi-fluid equilibria, the pressure-driven current component ($J_{\rm pres}$) is rigidly determined by the thermodynamic profiles and centrifugal forces. Consequently, scaling the total current by simply adjusting the pressure magnitude is physically inconsistent. Instead, we maintain a fixed target current $I_p$ by adjusting the poloidal current function $F(\psi_n)$ dynamically.

Using the parameterization in Eq.~(\ref{eq:F_function}), $F(\psi_n) = F_{\rm vac} [ 1 - \gamma_f (1 - \psi_n^{\beta_f})^{\alpha_f} ]$, the parameter $\gamma_f$ serves as the control variable. In each outer iteration, we calculate the integrated total current $I_{\rm total}^{(k)}$ and update $\gamma_f$ via a feedback rule:
\begin{equation}
    \gamma_f^{(k+1)} = \gamma_f^{(k)} + \delta \frac{I_{\rm target} - I_{\rm total}^{(k)}}{\partial I_{\rm total} / \partial \gamma_f},
\end{equation}
where $\delta$ is a damping factor and the derivative is estimated numerically. This approach ensures that the macroscopic engineering constraint ($I_p$) is satisfied while the internal multi-fluid force balance remains self-consistent.

The NFEQ code is benchmarked and compared with MHD rotation equilibrium solver as shown in \ref{sec:cmp}.

\section{Application to ENN Spherical Tokamaks}\label{sec:application}

To evaluate the impact of multi-fluid effects, we apply the model to two generations of p-$^{11}$B devices designed by the ENN Group: the experimental-scale EHL-2\cite{Liang2025} and the reactor-scale EHL-3B \cite{Liu2024}.

\subsection{Simulation Setup}

The input parameters for the two cases are summarized in Table~\ref{tab:params}. EHL-2 represents a high-field, high-temperature experiment, while EHL-3B is a conceptual reactor aimed at burning plasma conditions. For both cases, we assume a three-species plasma ($e, p, ^{11}\rm{B}$) and explore regimes with varying toroidal rotation frequencies. The last closed flux surface (LCFS) boundary shape is defined by $R=R_0+a\cos[\theta+\sin^{-1}(\delta)\cos\theta]$ and $Z=-\kappa a\sin\theta$, with $A=R_0/a$.

\begin{table}[h]
\caption{Design parameters for EHL-2 and EHL-3B. Profile parameters are set as $\alpha_{ns}=\beta_{ns}=\alpha_{Ts}=\beta_{Ts}=1.0$, $\alpha_{\Omega s}=2$, $\beta_{\Omega s}=4$, $\alpha_{f}=2$, $\beta_{f}=1$, and edge ratios are $T_{s,edge}/T_{s0}=N_{s,edge}/N_{s0}=0.05$. The density ratio is $N_{p0}/N_{B0}=9$.}
\label{tab:params}
\centering
\begin{tabular}{lccc}
\br
Parameter & Symbol & EHL-2 & EHL-3B \\
\mr
Major radius [m] & $R_0$ & 1.05 & 3.2 \\
Aspect ratio & $A$ & 1.85 & 1.7 \\
Elongation & $\kappa$ & 2.2 & 2.5 \\
Triangularity & $\delta$ & 0.5 & 0.6 \\
Toroidal field [T] & $B_{t0}$ & 3.0 & 5.0 \\
Plasma current [MA] & $I_p$ & 3.0 & 25.0 \\
Ion temp. [keV] & $T_{i0}$ & 30.0 & 150.0 \\
Electron temp. [keV] & $T_{e0}$ & 15.0 & 50.0 \\
Density [$10^{19} {\rm m}^{-3}$] & $N_{i0}$  & 10.0 & 25.0 \\
\br
\end{tabular}
\end{table}

\subsection{Physical Results and Discussion}

\subsubsection{Centrifugal Separation and Mach Number Thresholds}
The simulation results reveal that the deviation from single-fluid MHD is primarily governed by the boron Mach number $M_B$, or more accurately $M_B^2$. As demonstrated in our results (Figures \ref{fig:EHL2_w0}, \ref{fig:EHL2_w2}, \ref{fig:EHL2_w15}, \ref{fig:EHL3B_w0}, \ref{fig:EHL3B_w2}, \ref{fig:EHL3B_w5}), for $M_B < 0.5$, the centrifugal force is insufficient to overcome the thermal pressure, and the species remain nearly uniform on magnetic surfaces. This explains why single-fluid models often suffice for low-rotation experiments.

However, for $M_B > 1$, significant centrifugal separation occurs. In the high-rotation EHL-2 regime (Fig.~\ref{fig:EHL2_w15}) and EHL-3B regime (Fig.~\ref{fig:EHL3B_w5}), where $M_B \gtrsim 2$ at the outboard midplane, the heavy $^{11}\rm{B}$ ions are strongly shifted toward the low-field side (LFS), forming a "crescent" density distribution. This separation leads to a spatial mismatch between the fuel ions ($p$ and $^{11}\rm{B}$), which implies that the local fusion power density $P_f \propto n_p n_B \langle \sigma v \rangle$ may be reduced in the core and shifted outward compared to single-fluid predictions.

\subsubsection{Electrostatic Potential and Safety Factor}
The separation of ion species generates a self-consistent macroscopic electrostatic potential $\Phi$. In the reactor-scale EHL-3B, $\Phi$ reaches magnitudes of approximately 10--15 kV. This potential structure is non-uniform on flux surfaces and creates a poloidal electric field. Such a significant potential barrier can modify the radial electric field $E_r$, which is a critical factor for the suppression of micro-turbulence via $E \times B$ shear.

Furthermore, the safety factor $q(\psi)$ is modified by the centrifugal redistribution of the pressure gradient. In the reactor-relevant EHL-3B case, the extreme Shafranov shift and the outboard-peaked current density result in a $q$-profile with a very low central safety factor and high edge shear. These features are fundamentally linked to the multi-fluid nature of the plasma and provide a consistent equilibrium target for future stability analysis.

\subsubsection{Necessary for p-11B Modeling}
The application to EHL-2 and EHL-3B highlights that multi-fluid modeling is not merely a theoretical refinement but a practical necessity for p-$^{11}$B ST design. The spatial separation of species and the generation of multi-kV potentials are dominant features in the reactor regime that single-fluid models fail to resolve. This reduced model successfully establishes the baseline equilibrium required for subsequent investigations into p-$^{11}$B stability and transport\cite{Hinton1985,Abel2013}.

\subsubsection{Scaling of Mach Number in Reactors}
It is noteworthy that as p-$^{11}$B devices scale from experimental size (EHL-2) to reactor size (EHL-3B), the achievable toroidal Mach number may theoretically decrease due to the substantial increase in ion thermal velocity ($v_{th} \propto \sqrt{T_i}$) at ignition temperatures. However, the multi-fluid effects remain a first-order concern for two reasons. First, due to the mass disparity, even a subsonic proton flow ($M_p \sim 0.3$) results in a transonic boron flow ($M_B \sim 1.0$). Second, the absolute centrifugal potential energy scales as $R^2$. Thus, even at reduced angular frequencies, the large major radius of EHL-3B ensures that the spatial species separation and the resulting multi-kV electrostatic potentials remain significant.

\section{Conclusions and Discussion}\label{sec:summary}

In this work, we have developed a reduced multi-fluid equilibrium model tailored for the investigation of strongly rotating proton-boron plasmas in spherical tokamaks. By prioritizing the dominant physical effects---toroidal centrifugal forces and the resulting self-consistent electrostatic potentials---while strategically neglecting poloidal flow inertia and pressure anisotropy, we have established a computationally robust and physically insightful theoretical baseline. 

The primary conclusions drawn from the application of this model to the ENN EHL configurations are:
\begin{enumerate}
    \item \textbf{Mach Number Threshold:} The ion Mach number is the critical parameter. Multi-fluid effects are negligible for $M < 0.5$ but become dominant for $M > 1$.
    \item \textbf{Species Separation:} In the reactor regime, boron ions exhibit a pronounced LFS accumulation, leading to a fuel density mismatch that single-fluid models cannot resolve.
    \item \textbf{Macroscopic Potential:} A self-consistent potential on the order of 10 kV is generated in EHL-3B to maintain quasi-neutrality, significantly influencing the internal electric field structure.
\end{enumerate}


Although this work focuses on the extreme parameter space of p-$^{11}$B spherical tokamaks, it is important to note that the reduced multi-fluid framework developed herein is generic and applicable to a wider range of fusion scenarios. 
\begin{itemize}
    \item \textbf{High-Z Impurity Transport:} In standard D-T tokamaks (e.g., ITER or DEMO), heavy impurities such as tungsten ($W$, $m_W \approx 184 m_p$) are subject to strong centrifugal forces even at low bulk plasma rotation ($M_{D-T} \ll 1$). This model can accurately predict the poloidal asymmetry of impurities, which is critical for understanding impurity accumulation and core radiative collapse.
    \item \textbf{Isotopic Separation:} For D-T or D-$^3$He plasmas, differential centrifugal drifts can lead to isotopic variation on flux surfaces. This solver provides a tool to assess fuel spatial mismatch and optimize fueling strategies in rotating burning plasmas.
    \item \textbf{Advanced Tokamak Regimes:} The capability to resolve the interplay between rotation, safety factor $q$, and potential $\Phi$ makes this tool valuable for analyzing internal transport barriers (ITBs) in strongly rotating Advanced Tokamak (AT) modes.
\end{itemize}
Thus, the solver serves as a versatile platform for investigating multi-species equilibrium physics across various magnetic confinement concepts.

As stated in the introduction, the present work focuses on establishing the theoretical foundation. Future investigations will extend this baseline to address:
\begin{itemize}
    \item \textbf{Transport and Fusion Power:} Integrating this equilibrium with transport codes to accurately predict the fusion gain $Q$ accounting for fuel mismatch.
    \item \textbf{Helium Ash Exhaust:} Investigating if the centrifugal separation mechanism can be exploited to naturally remove helium ash ($\alpha$-particles) to the LFS.
    \item \textbf{Stability:} Using the non-isobaric multi-fluid equilibrium as a basis for modified Mercier and ballooning stability studies.
    \item \textbf{Free-Boundary Reconstruction:} Extending the solver to free-boundary conditions for experimental control and reconstruction.
\end{itemize}

In conclusion, the reduced multi-fluid model presented here is an essential tool for the accurate physical description of p-$^{11}$B spherical tokamak reactors, establishing a robust starting point for the integrated modeling effort required to realize aneutronic fusion energy.

\begin{figure*}
\centering
\includegraphics[width=17.5cm]{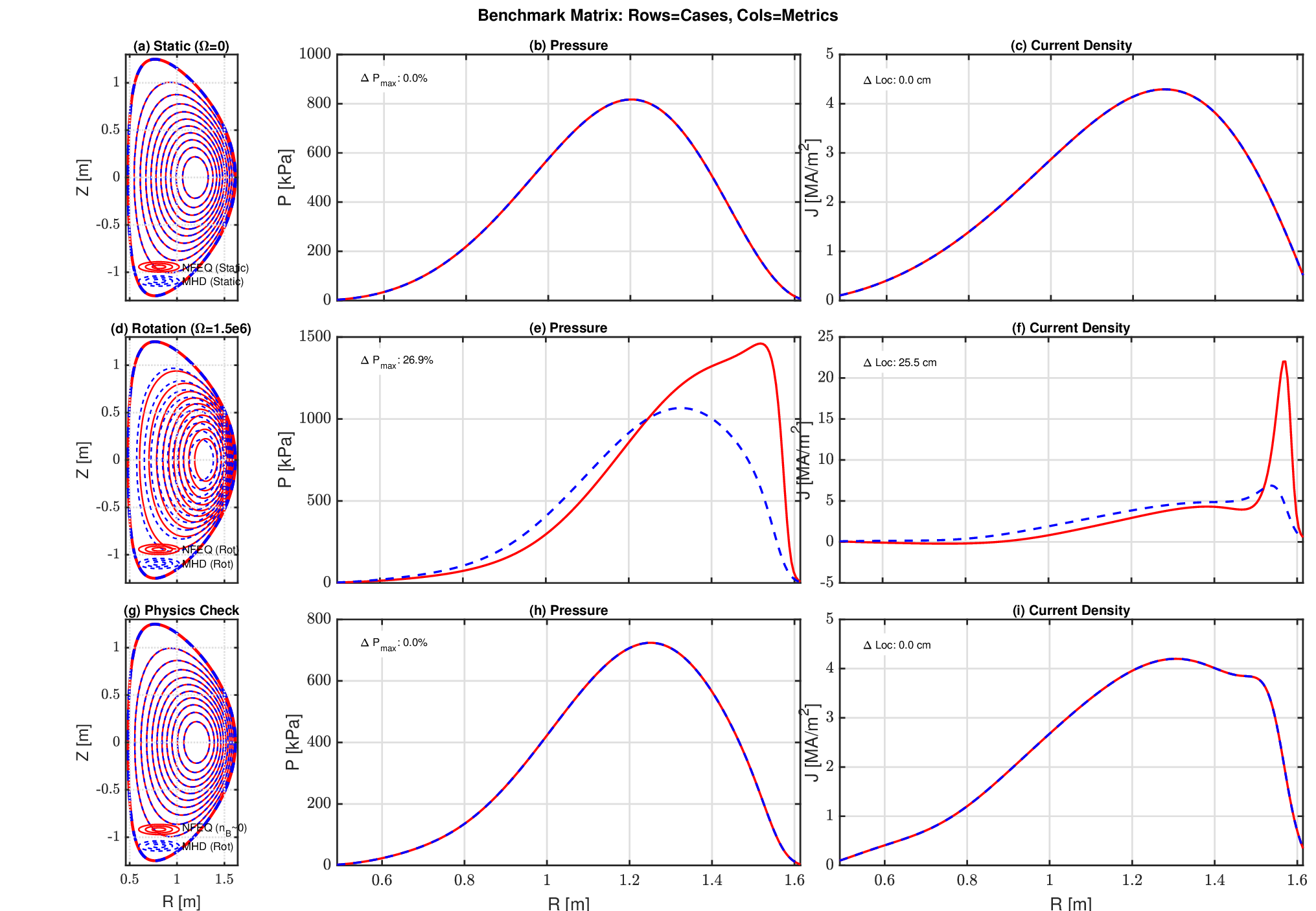}
\caption{Benchmark of the multi-fluid equilibrium model against a standard MHD solver. (a)-(c) Static case ($M=0$) showing perfect agreement. (d)-(f) Rotating case ($M \neq 0$) with $p+^{11}$B, highlighting significant species separation in the multi-fluid result. (g)-(i) Rotating case with negligible boron ($n_B \simeq 0$), where the multi-fluid model correctly reduces to the single-fluid solution.}
\label{fig:nfeq_mhd_cmp}
\end{figure*}

\ack
The authors would like to thank the ENN Group for their support of this research.

\appendix
\section{Verification against MHD Equilibrium}\label{sec:cmp}

To validate the numerical implementation, we benchmark the multi-fluid solver against a standard single-fluid MHD rotation solver \cite{Li2026}. As shown in Figure \ref{fig:nfeq_mhd_cmp}, in the limit where rotation vanishes ($M=0$) or the impurity concentration is zero ($n_B=0$), the multi-fluid model yields results identical to the MHD model. However, when $M \neq 0$ and multiple species are present ($n_p, n_B \neq 0$), the multi-fluid species separation effect becomes significant, demonstrating the capability of the new code to capture physics beyond the single-fluid description.

\end{document}